\documentclass{article}

\usepackage{graphicx}
\usepackage[latin1]{inputenc}
\usepackage[english]{babel}
\usepackage{multicol}

\usepackage{amssymb}
\usepackage{amsmath}
\usepackage{lineno}
\usepackage{fancyhdr}
\usepackage{tabulary}
\usepackage{times}
\usepackage{latexsym}

\title{\textbf{THE COSMOLOGICAL CONSTANT PROBLEM FROM THE POINT OF VIEW OF STRING THEORY}} 
\author{Nicolás Avilán V.\footnote{ng.avilan47@uniandes.edu.co}, Jos\'e Rolando Rold\'an.\\Universidad de Los Andes\\Bogotá-Colombia}

\begin{document}
\maketitle

\begin{center}
\begin{minipage}{11.5cm}

\centerline{\bf\small ABSTRACT}
\vspace{-0.05cm} \noindent{\footnotesize 
The quantum field theory prediction of the cosmological constant is 120 orders of magnitude higher than the observed value. This is known as the cosmological constant problem. Here, we deal with the cosmological constant as a scalar field generated by the reduced extra dimensions, following Kaluza-Klein reduction, in the superstring theory frame, where the vacuum energy is the minimum of this field. We find methods for calculating the cosmological constant, whose value is much lower than those obtained with the current methods of quantum field theory. We conclude that the small cosmological constant value could be explained by the behavior of the mentioned scalar field, and finally we discuss a possible explanation for the observed accelerated expansion of the Universe.  \vspace{0.15cm}

{\bf Keywords:} Cosmological constant problem, superstring theory, Kaluza-Klein reduction.}
\end{minipage}
\end{center}\vspace{0.3cm}

\section{Introduction}
The cosmological constant problem is considered to be in the frontier of particle physics and cosmology. This problem could be deeply connected with the difficulties to unify general relativity and quantum field theory. Since the observation of the accelerated expansion of the Universe, the cosmological constant problem recovered a leading place in theoretical physics \cite{Padma}.\\
In this work we follow an approach to string theory considering general features that could be evidence of strings as a fundamental structure in nature. In fact, fundamental constants, like the cosmological constant $\Lambda$\footnote{There are several works considering similar analysis for the fine structure constant $\alpha$ becomign a scalar field\cite{Abou}.}, could become fields; giving place to the existence of higher dimensions, generating scalar fields by compactification of extra dimensions, and the possible existence of scalar fields coming from $D$-brane contributions to the action. These fields could be an indirect evidence for strings.\\
In this work we obtain an effective cosmological constant and analyze the possibilities it having a tiny positive value as suggested by the current cosmological observations. Here the cosmological constant has mainly two contributions, a bulk space one, coming from extra dimensional reduction, and becoming a scalar field in a 4-dimensional world, and a $D$-brane contribution through the Dirac-Born-Infeld action. The methods developed in this work could be a hint to solve the cosmological constant problem. \\
This paper is organized as follows. In the second section we present the cosmological constant problem, the third section is devoted to calculate the contributions from bulk space and $D$-brane, there we will consider the implications of the possible bulk cosmological constant values; finally we discuss the results and present conclusions.\\
%
%
\section{The Cosmological Constant Problem}
Currently the cosmological constant problem points to understand in a natural way why the vacuum energy density $\rho_v$ is so small; this problem is formulated today to investigate why $\rho_v$ is not only small, but also, comparable to the present mass density of the Universe\cite{Weinberg89, Weinberg2000}.
The vacuum energy density expressed by the cosmological constant, observationally appears to have a small positive value.\\
Although the cosmological constant was first proposed by Einstein as a geometrical term in the general relativity field equations, since Zeldovich's work\cite{Zel} it is not considered as pure geometrical contribution any longer. Now the cosmological constant can be interpreted as the ground state of the field, the vacuum, and it must have gravitational observable effects in the Universe dynamics. For this reason the cosmological constant is written in Einstein's field equations at the right side, as a common energy source.
\paragraph{Vacuum energy density.}
The formulation of a free quantum field theory needs an infinite number of quantum simple harmonic oscillators. The contributions to the vacuum energy density coming from each oscillation mode are given by
\begin{equation}
\langle\rho\rangle=\int_0^\infty \frac{d^3\vec k}{(2\pi)^3}\frac1{2}\sqrt{k^2+m^2}
\end{equation}
If we take all oscillation modes the integral diverges, but if we consider a cutoff $k_{max}$ for contributions in a specific mode and also that the momentum contribution is greater than the mass one, the density energy becomes 
\begin{equation}
\langle\rho\rangle=\int_0^{k_{max}} \frac{4\pi k^2dk}{(2\pi)^3}\frac1{2}\sqrt{k^2+m^2}\approx \frac1{(2\pi)^2}\int_0^{k_{max}}k^3dk=\frac{k_{max}^4}{16\pi^2} 
\end{equation}
Commonly the cutoff is defined for UV frecuencies. Because of this result the energy density is proportional to $M^4$, being $M$ the mass scale of the considered interaction. For Quantum gravity $M=M_{Pl}=(8\pi G)^{-1/2}\sim10^{18}GeV$ and
\begin{equation}
\rho_\Lambda^{Pl}\sim(10^{18}GeV)^4\sim2\times10^{110}erg/cm^3
\end{equation}
whereas cosmological observations, as the accelerated expansion of the Universe\cite{Padma,Carroll}, constrain this quantity to be
\begin{equation}
\vert \rho_\Lambda^{Obs}\vert\leq(10^{-12}GeV)^4\sim2\times10^{-10}erg/cm^3
\end{equation}
Therefore
\begin{equation}
\frac{\rho_\Lambda^{Pl}}{\rho_\Lambda^{Obs}}\sim 10^{120}
\end{equation}
which is an evidence of the lack understanding of the fundamental interactions.

%
%
\section{Contributions to the Cosmological Constant}
The action at large distances, in the presence of a $D$-brane, has the form
\begin{equation}
S=S_{Bulk}+S_{D-brane}
\end{equation}
In string theory, $S_{Bulk}$ is the bosonic part of the appropriate low-energy supergravity theory, while the action $S_{D-brane}$ describes the effective dynamics of the massless bosonic modes on the $D$-brane world volume\cite{Abou}. In our analysis we consider $S_{Bulk}=S_g$, the gravity contribution in a $d$-dimensional space-time with a cosmological constant. The $D$-brane contribution is given by the Dirac-Born-Infeld (DBI) action, i.e., $S_{D-brane}=S_{DBI}$.\\
The gravitational action in the $d$-dimensional space-time is given by
\begin{equation}
S_g=\frac1{16\pi G_{d}}\int d^{d}x\sqrt{-\det g_{MN}}(R_d-2\Lambda_d)
\end{equation}
where $G_{d}$ is the gravitational constant in $d$ dimensions, the indices with capital letters take values $M,N=0,1,2..,d-1$; $R_d$ is Ricci scalar and $\Lambda_d$ is the cosmological constant of the bulk space.\\
The $D$-brane contribution is given by the Dirac-Born-Infeld (DBI) action.
For a $D_3$-brane the DBI action is
\begin{equation}
S_{DBI}=-T\int d^{4}x\sqrt{-\det (g_{\mu\nu}+kF_{\mu\nu})}
\end{equation}
where $k$ has the constant value $2\pi\alpha'$ and $T$ is the tension of the $D$-brane. $g_{\mu\nu}$ is the metric tensor on the $D$-brane, and $F_{\mu\nu}$ is the strength of the field such that the expansion in powers of $F$ gives the Maxwell action at the second order in $k$. In this paper greek letters take values $\mu,\nu, \alpha,\beta=0,1,2,3$.
%
%
\paragraph{Bulk space contribution.} We consider the bulk space as a $d=4+D$ space-time, where each of the $D$ extra dimension is compacted in a circle of size $2\pi R$. To obtain the contributions of the bulk space to the $4$-dimensional world we perform a Kaluza-Klein reduction, considering the special case when the metric elements depend only on the four world coordinates, i.e., $g_{MN}(x^\alpha)$. We need to separate the space-time terms in the Ricci scalar calculation as follows
\begin{equation}
R_d=g^{AB}{R^N}_{ANB}=\underbrace{g^{\mu\nu}{R^\alpha}_{\mu\alpha\nu}}_{R_{4}}+g^{\alpha\beta}{R^n}_{\alpha n\beta}+g^{mn}{R^N}_{mNn}+2g^{m\alpha}{R^N}_{mN\alpha}
\end{equation}
Where $m,n,o,p,...=4,5,..., d-1$. Taking into account these geometric considerations we can obtain the usual Ricci scalar $R_{4}$ for $4$-dimensional space-time and a set of scalar and vector fields coming from the reduction of extra dimensions of metric elements. Developing these terms we obtain
\begin{equation}
R_d=R_{4}+4\partial_\alpha \phi\partial^\alpha\phi-\frac1{4}g^{mn}g^{op}\partial_\alpha g_{mp}\partial^\alpha g_{no}-\frac1{4}g^{mn}F_{m\delta\alpha}F^{\alpha\delta}_n
\end{equation}
Where 
\begin{equation}
\phi=-\frac1{4}ln(Det g_{mn})\quad and \quad F_{n\alpha\beta}=g_{n\alpha,\beta}-g_{n\beta,\alpha}
\end{equation}
After this reduction the gravitational action takes the form
\begin{equation}
S_g=\frac{1}{16\pi G_d}\int d^{d}x\sqrt{-\det g_{MN}}(R_{4}+4\partial_\alpha \phi\partial^\alpha\phi-\frac1{4}g^{mn}g^{op}\partial_\alpha g_{mp}\partial^\alpha g_{no}-\frac1{4}g^{mn}F_{m\delta\alpha}F^{\alpha\delta}_n-2\Lambda_d)
\end{equation}
As we now have each element in the $4$-dimensional space-time, we must integrate over the extra dimensions, giving\footnote{To obtain the exponential factor we used the Bailin and Love ansatz \cite{Love}.}
\begin{equation}
S_g=\frac{(2\pi R)^{D}}{16\pi G_d}\int d^{4}x\sqrt{-\det g_{\mu\nu}}e^{-2\phi}(R_4-2\Lambda_d+4\partial_\alpha \phi\partial^\alpha\phi-\frac1{4}g^{mn}g^{op}\partial_\alpha g_{mp}\partial^\alpha g_{no}-\frac1{4}g^{mn}F_{m\delta\alpha}F^{\alpha\delta}_n)
\end{equation}
Here we see the known relation between the gravitational constant and the extra dimensions
\begin{equation}
G_d=G_{4+D}=(2\pi R)^DG_4
\end{equation}
%
%
\paragraph{$D$-brane contribution.} The square root in the action $(8)$ makes the work difficult. Using matrix notation, let $M=kg^{-1}F$ and because 
\begin{equation}
\det(g_{\mu\nu}+kF_{\mu\nu})=\det(g_{\mu\nu}+kF_{\mu\nu})^T=\det(g_{\mu\nu}-kF_{\mu\nu})
\end{equation}
then the root in DBI action becomes
\begin{equation}
\sqrt{-\det(g+kF)}=\sqrt{-\det g}\sqrt{\det(1+M)}=\sqrt{-\det g}\bigl[\det(1-M^2)\bigr]^{1/4}
\end{equation}
after some matrix analysis we arrived at an expansion of the action in powers of $kF$\cite{BBS}. Thus, the action up to second order in $k$ is 
\begin{equation}
S_{DBI}=-T\int d^{4}x\sqrt{-\det (g_{\mu\nu})}\left(1+\frac1{4}k^2F_{\mu\nu}F^{\mu\nu}-...\right)
\end{equation}
The total action $(6)$, including Bulk and $D$-brane contributions, takes the form
\begin{multline}
S=\frac1{16\pi G_4}\int d^{4}x\sqrt{-\det (g_{\mu\nu})}e^{-2\phi}\biggl[R_{4}-2\Lambda_d-16\pi G_4T\left(1+\frac1{4}k^2F_{\mu\nu}F^{\mu\nu}\right)\\+\biggl(4\partial_\alpha \phi\partial^\alpha\phi-\frac1{4}g^{mn}g^{op}\partial_\alpha g_{mp}\partial^\alpha g_{no}-\frac1{4}g^{mn}F_{m\delta\alpha}F^{\alpha\delta}_n\biggr)\biggr]
\end{multline}
Now we apply a Weyl transformation to the action $(19)$, making a field redefinition
\begin{equation}
\tilde g_{\mu\nu}=\exp\left(-2\phi\right)g_{\mu\nu}
\end{equation}
After applying this transformation on the metric components in the volume element and the Ricci scalar, we have
\begin{multline}
S'_g=\frac1{16\pi G_{4}}\int d^{4}x\sqrt{-\det \tilde g_{\mu\nu}}\biggl[\tilde R_{4}-2\Lambda_d\exp\left(2\phi\right)-16\pi G_4T\exp(4\phi)-\\4\pi G_4T\exp(4\phi)k^2F_{\mu\nu}F^{\mu\nu} + \exp\left(2\phi\right)\biggl(4\partial_\alpha \phi\partial^\alpha\phi-\frac1{4}g^{mn}g^{op}\partial_\alpha g_{mp}\partial^\alpha g_{no}- \frac1{4}g^{mn}F_{m\delta\alpha}F^{\alpha\delta}_n\biggr)\biggr]\\
\end{multline} 
where $\tilde g_{\mu\nu}$ and $\tilde R_4$ are metric components and Ricci scalar, respectively, after the Weyl conformal transformation.
%
%
\paragraph{Effective cosmological constant.} Comparing equations $(7)$ and $(21)$ we find that the effective cosmological constant in $4$-dimensional space-time that is given by\footnote{Kamani obtained general results in a $(p+1)$ dimensional world\cite{Kamani}, such that the ones we obtain here are a particular case for $p=3$.}
\begin{equation}
\Lambda_{eff}=\Lambda_d\exp\left(2\phi\right)+\frac{8\pi G_d}{(2\pi R)^D}T\exp\left(4\phi\right)
\end{equation}
Here we can see that compactification influences the cosmological constant value and it depends on $\Lambda_d$, $R$, $T$ and the scalar field $\phi$. Adjusting these parameters we can obtain a small cosmological constant.
%
%
\subsection{Cosmological constant for $\Lambda_d=0$}
Considering the bulk space without cosmological constant, $\Lambda_d=0$, there are only contributions from the $D_3$-Brane, i.e.
\begin{equation}
\Lambda_{eff}=\Lambda_{D-brane}=\frac{8\pi G_d}{(2\pi R)^D}T\exp\left(4\phi\right)
\end{equation}
We have a small $\Lambda_{eff}$ if $\phi\rightarrow -\infty$. In this case we don't have bulk space contributions.
%
%
\subsection{Cosmological constant for $\Lambda_d>0$}
By direct inspection we can see that $\Lambda_d>0$ in equation $(22)$ gives us always a positive $\Lambda_{eff}$ and the scalar field can take values $-\infty<\phi<\infty$. A small and positive cosmological constant can be obtained if $\Lambda_d$ is small and positive, if $R$ is large or if $\phi\rightarrow -\infty$.\\
Considering the $D_3$-branes of the superstring theory in $d=10$, with $8\pi G_{10}^2=\frac1{2}(2\pi)^7g_s^2\alpha'^4$ and $T=\frac1{8\pi^3\alpha'^2}$, the effective cosmological constant is
\begin{equation}
\Lambda_{eff}=\Lambda_{10}\exp\left(2\phi\right)+\frac{g_s\alpha'^2}{8\pi^2R^6}\exp\left(4\phi\right)
\end{equation}
This is the cosmological constant corresponding to the type $IIB$ superstring theory for $D_3$-branes.\\
We can suppose that there exists a specific value $\phi=\phi_0$ that allows us to obtain the observed cosmological constant value. In this situation we have several parameters that by adjustments give the desired value for $\Lambda_{eff}$; these are $\Lambda_{10}$, $R$, $\phi_0$ and $g_s$. In this way we can describe a world with small and positive cosmological constant if $\phi_0\rightarrow -\infty$ and we have finite values for $\Lambda_{10}$ $R$ and $g_s$; or small $\Lambda_{10}$ and large $R$ for $\phi_0\rightarrow 0$. In this way the behavior of $\phi$ governs the mechanism for adjusting the cosmological constant.
%
%
\subsection{Cosmological constant for $\Lambda_d<0$}

If we consider a negative cosmological constant $\Lambda_d$, there is a minimum for $\Lambda_{eff}$ when $\phi=\phi_\varnothing$ with
\begin{equation}
e^{2\phi_\varnothing}=\frac{-\Lambda_d}{16\pi G_4T}
\end{equation}
where the effective cosmological constant takes the value
\begin{equation}
\Lambda_{eff}(\phi_\varnothing)=\frac1{2}\Lambda_d\exp \left(2\phi_\varnothing\right)
\end{equation}
We have positive and negative values for $\Lambda_{eff}$ when $-\infty<\phi<\infty$ with $\Lambda_d<0$, therefore the minimum of $\Lambda_{eff}$ is negative too.\\
If we study again the case of a $D_3$-brane of the type IIB superstring theory, the cosmological constant is given again by equation $(22)$, but here we have $\Lambda_d<0$ rather than $\Lambda_d>0$.\\
Let $\phi_*$ be such that $\Lambda_{eff}(\phi_*)=0$. We could have a tiny positive cosmological constant for $\phi$ a little bit bigger than $\phi_*$, but this is an unstable state; we would also expect a behavior in which $\phi$ rolls down to $\phi_*$ and then to $\phi_\varnothing$ where $\Lambda_{eff}$ establishes itself with a value
\begin{equation}
\Lambda_{eff}(\phi_\varnothing)=-\frac{g_s\alpha'^2}{8\pi^2R^6}e^{4\phi_\varnothing}
\end{equation}
This would be related with a future phase of decelerated expansion of the Universe, see figure 1.\\
\begin{figure*}
    \centering
    \includegraphics[width=0.5\textwidth,angle=0]{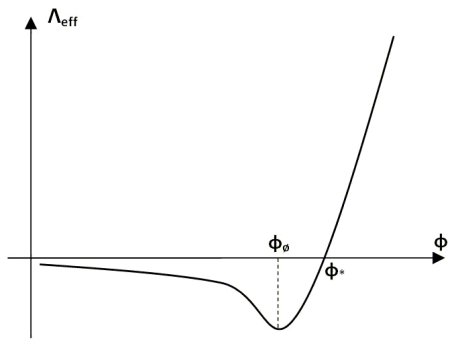}
    \caption{Dependence of $\Lambda_{eff}$ vs. $\phi$, for $\Lambda_d<0$.}
    \label{figure}
\end{figure*}

This value is bound by the anthropic argument, it constrains us to discard large values for the cosmological constant both positive and negative ones, because with a large negative value the universe colapses rapidly, while with a large positive one all matter in the Universe disperses away\cite{WeinbergB}.\\

%
%
\section{Discussion and Conclusions}
It is an interesting possibility to detect the massive scalar field $\phi$, produced by compactification of extra dimensions, as a possible indirect evidence of string theory. It is a very important approach we put forward explicitly, because today it is not possible to develop any experimental setup for a direct confirmation of string theory.\\ 
The effective cosmological constant was built by contributions of a bulk space coupled with a $D$-brane; it was possible to choose the parameters behavior such that there was a predicted cosmological constant closer to the observed one. In all the three cases considered for $\Lambda_d$, lowering the cosmological constant was possible; in the case $\Lambda_d<0$ there is a negative ground state energy related with a decelerated expansion.\\ 
Although it is possible to tune the parameters of the model, the calculations developed lack physical criteria to constrain each one of them.\\
%
%
\section{Acknowledgements}
This work and our active participation in the ``XIII Congreso Nacional del F\'{i}sica"  have been supported by ``Comité de Investigaciones y Postgrados de la Facultad de Ciencias de la Universidad de Los Andes".
%
%

\end{document}